\documentclass{article}
\usepackage{fullpage}
\usepackage[american]{babel}
\usepackage{amssymb}
\usepackage{amsmath}
\usepackage{epsfig}
\usepackage{slashbox}
\usepackage{pseudocode}
\usepackage[geometry]{ifsym}
\usepackage{multirow}

\newcommand{\eqn}[1]{(\ref{#1})}
\newcommand{\beql}[1]{\begin{equation}\label{#1}}
\newcommand{\eeq}{\end{equation}}
\newtheorem{theo}{Theorem}
\newtheorem{lemma}{Lemma}

\title{Complete Solutions for a Combinatorial Puzzle in Linear Time}
\author{Lei Wang,Xiaodong Wang,Yingjie Wu, and Daxin Zhu}

\begin{document}
\maketitle

\begin{abstract}
In this paper we study a single player game consisting of $n$ black checkers and $m$ white checkers, called shifting the checkers. We have proved that the minimum number of steps needed to play the game for general $n$ and $m$ is $nm + n + m$.
We have also presented an optimal algorithm to generate an optimal move sequence of the game consisting of $n$ black checkers and $m$ white checkers, and finally, we present an explicit solution for the general game.
\end{abstract}

\section{Introduction}

Combinatorial games often lead to interesting, clean problems in algorithms and complexity theory. Many classic games are known to be computationally intractable. Solving a puzzle is often a challenge task like solving a research problem. You must have a right cleverness to see the problem from a right angle, and then apply that idea carefully until a solution is found.

In this paper we study a single player game called shifting the checkers. The game is similar to the Moving Coins puzzle [2,3,7], which is played by re-arranging one configuration of unit disks in the plane into another configuration by a sequence of moves, each repositioning a coin in an empty position that touches at least two other coins. In our shifting checkers game, there are $n$ black checkers and $m$ white checkers put on a table from left to right in a row. The $n+m+1$ positions of the row are numbered $1,\cdots,n+m+1$.
Initially, the $n$ black checkers are put in the position $1,\cdots,n$, and the $m$ white checkers are put in the position $n+2,\cdots,n+m+1$. The position $n+1$ is initially vacant.
In the final state of the game, the left most $m$ positions numbered $1,\cdots, m$ are occupied by white checkers, and the right most $n$ positions numbered $m+2,\cdots, m+n+1$ are occupied by black checkers, leaving the position $m+1$ vacant, as shown in Fig. 1.

\begin{figure}
\centering
\includegraphics[width=7.5cm,height=3.5cm]{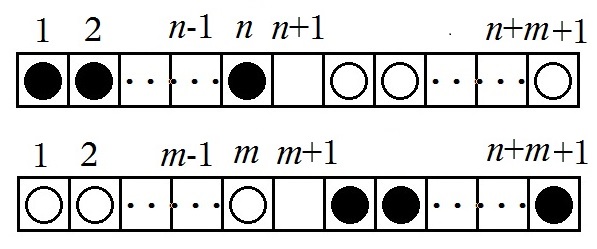}
\caption{Shifting checkers}
\end{figure}

There are only two permissible types of moves. A move of the game consists of sliding one checker into the current vacant position, or jumping over the adjacent checker into the current vacant position.
The goal of the game is to make a small number of moves to reach the final state of the game.

We are interested in algorithms which, given integers $n$ and $m$, generate the corresponding move sequences to reach the final state of the game with the smallest number of steps.
In this paper we present an optimal algorithm to generate all of the optimal move sequences of the game consisting of $n$ black checkers and $m$ white checkers.

This paper is structured as follows.

In the following 4 sections we describe the algorithms and our computing experience with the algorithms for generating optimal move sequence of the general game consisting of $n$ black checkers and $m$ white checkers.
In Section 2 we describe a new variant tree search based algorithm for generating all optimal solutions for the shifting checkers games of the small size.
A linear time recursive construction algorithm is proposed in Section 3.
Based on the recursive algorithm proposed in Section 3, an explicit solution for the optimal move sequence of the general game is presented in Section 4. we discuss the number of optimal solutions of the game in Section 5.
Some concluding remarks are in Section 6.

\section{A Backtracking Algorithm}
In a row of checkers of the game, if two checkers have different colors and the black checker is on the left of the white checker, then the two checkers are called an inversion pair.
For example, in the initial state of the game consisting of $n$ black checkers and $m$ white checkers, since all of the $n$ black checkers are on the left of all the $m$ white checkers, there is total $nm$ inversion pairs. On the other hand, in the final state of the game, since all of the $n$ black checkers are on the right of all the $m$ white checkers, there are no inversion pairs in this case.

Similarly, for the vacant position, if a black checker is on the left of the vacant position, or a white checker is on the right of the vacant position, then the checker and the vacant position are called a vacant inversion pair.
For example, in the initial state of the game, since all of the $n$ black checkers are on the left of the vacant position, and all of the $m$ white checkers are on the right of the vacant position, there are total $n+m$ vacant inversion pairs. On the other hand, in the final state of the game, since all of the $n$ black checkers are on the right of the vacant position, and all of the $m$ white checkers are on the left of the vacant position, there are no vacant inversion pairs in this case.

Of the two types of checker moves, we can further list 12 different cases of the moves into a table, as shown in Table 1.
Sliding a black checker right into the current vacant position is denoted as $slide(b,r)$. The other three moves $slide(b,l)$, $slide(w,r)$, and $slide(w,l)$ are defined similarly.
Jumping a black checker right over the adjacent white checker into the current vacant position is denoted as $jump(b,w,r)$.
The other 7 moves $jump(b,w,l)$, $jump(w,b,r)$, $jump(w,b,l)$, $jump(b,b,r)$, $jump(b,b,l)$, $jump(w,w,r)$, and $jump(w,w,l)$ are defined similarly.
These 12 cases of moves are numbered from 1 to 12.

The column \textit{Inversions} of Table 1 denote the inversion increment of the checker row when the corresponding case of moves applied. Similarly, the column \textit{V-Inversions} of Table 1 denotes the vacant inversion increment of the checker row when the corresponding case of moves applied.

It is not difficult to verify the following facts on the optimal solutions to play the game.

\begin{lemma}\label{le1}
Any optimal solution for playing the game of shifting the checkers with minimum number of moves consists of only the classes of moves numbered from 1 to 4 in Table 1.
\end{lemma}
\noindent{\bf Proof.}
We first notice that in an optimal solution, when jumping, a checker may only jump a single checker of the opposite color. If a checkers jump over the adjacent checker of the same color into the vacant position, then we will arrive an unfavorable status. For example, if a step of $jump(b,b,r)$ is applied, then the white checkers located in the right of the vacant position will be stuck unless a step of $slide(b,l)$ is applied immediately. But these two steps can be substituted by only one step $slide(b,r)$. The other cases can be analyzed similarly. Therefore, we know that the classes of moves numbered from 9 to 12 in Table 1 will not appear in an optimal solution with the minimum number of moves.

We have known that from the initial state of the game to the final state of the game, there are total of $nm$ inversions and $n+m$ vacant inversions to be reduced. From Table 1 we see that for each step of moves numbered from 1 to 8 in, at most 1 inversion or vacant inversion can be reduced. Therefore it requires at least $nm+n+m$ steps to play the game consisting of $n$ black checkers and $m$ white checkers. In other words, $nm+n+m$ is a lower bound for solving the game. In the next section, we will present an optimal solution for the game in exactly $nm+n+m$ steps.
If a solution for the game contains any steps of moves numbered from 5 to 8 in Table 1, then these steps will increase the inversions or the vacant inversions of the checkerboard, and thus the number of steps to play the game must be no less than $nm+n+m+2$. Therefor, a solution for the game containing any steps of moves numbered from 5 to 8 in Table 1 cannot be an optimal solution of the game.

Summing up, an optimal solution for playing the game of shifting the checkers with minimum number of moves consists of only the steps of moves numbered from 1 to 4 in Table 1.
$\Box$

\begin{table}
\caption{All cases of checker moves}
\begin{center}
\begin{tabular}{|c|c|c|c|c|}
\hline
\textit{No.} & \textit{Move} & \textit{Change} & \textit{Inversions} & \textit{V-Inversions} \\\hline
1 & $slide(b,r)$
&\raisebox{-.6ex}{\FilledCircle \ \Square} $\rightarrow$ \raisebox{-.6ex}{\Square \ \FilledCircle}
& 0 & -1\\\hline

2 & $slide(w,l)$
&\raisebox{-.6ex}{\Square \ \Circle} $\rightarrow$ \raisebox{-.6ex}{\Circle \ \Square}
& 0 & -1\\\hline

3 & $jump(b,w,r)$
&\raisebox{-.6ex}{\FilledCircle \ \Circle \ \Square} $\rightarrow$ \raisebox{-.6ex}{\Square \ \FilledCircle \ \Circle }
& -1 & 0\\\hline

4 & $jump(b,w,l)$
&\raisebox{-.6ex}{\Square \ \FilledCircle \ \Circle} $\rightarrow$ \raisebox{-.6ex}{\Circle \ \FilledCircle \ \Square}
& -1 & 0\\\hline

5 & $jump(w,b,r)$
&\raisebox{-.6ex}{\Circle \ \FilledCircle \ \Square} $\rightarrow$ \raisebox{-.6ex}{\Square \ \FilledCircle \ \Circle }
& 1 & 0\\\hline

6 & $jump(w,b,l)$
&\raisebox{-.6ex}{\Square \ \Circle \ \FilledCircle} $\rightarrow$ \raisebox{-.6ex}{\FilledCircle \ \Circle\  \Square}
& 1 & 0\\\hline

7 & $slide(b,l)$
&\raisebox{-.6ex}{\Square \ \FilledCircle} $\rightarrow$ \raisebox{-.6ex}{\FilledCircle \ \Square}
& 0 & 1\\\hline

8 & $slide(w,r)$
&\raisebox{-.6ex}{\Circle \ \Square} $\rightarrow$ \raisebox{-.6ex}{\Square \ \Circle }
& 0 & 1\\\hline

9 & $jump(w,w,r)$
&\raisebox{-.6ex}{\Circle \ \Circle \ \Square} $\rightarrow$ \raisebox{-.6ex}{\Square \ \Circle \ \Circle }
& 0 & 2\\\hline

10 & $jump(b,b,l)$
&\raisebox{-.6ex}{\Square \ \FilledCircle \ \FilledCircle} $\rightarrow$ \raisebox{-.6ex}{\FilledCircle \ \FilledCircle \ \Square}
& 0 & 2\\\hline

11 & $jump(w,w,l)$
&\raisebox{-.6ex}{\Square \ \Circle \ \Circle} $\rightarrow$ \raisebox{-.6ex}{\Circle \ \Circle \ \Square}
& 0 & -2\\\hline

12 & $jump(b,b,r)$
&\raisebox{-.6ex}{\FilledCircle \ \FilledCircle \ \Square} $\rightarrow$ \raisebox{-.6ex}{\Square \ \FilledCircle \ \FilledCircle }
& 0 & -2\\
\hline
\end{tabular}
\end{center}
\end{table}

From Lemma~\ref{le1}, we can conclude that the following theorem holds.

\begin{theo}\label{th1}
For the general game of shifting the checkers consisting of $n$ black checkers and $m$ white checkers, it needs at least $nm+n+m$ steps to reach the final state of the game from its initial state.
\end{theo}

According to Theorem~\ref{th1}, if we can find a move sequence to reach the final state of the game with $nm+n+m$ steps, then the sequence will be an optimal move sequence, since no move sequence can reach the final state of the game in less than $nm+n+m$ steps. In order to study the structures of the optimal solutions for the general game of shifting the checkers, we first present a backtracking algorithm [1,5,6] to generate all optimal solutions of the games with small size.

In the algorithm described above, the parameter $i$ is the current number of steps and the parameter $e$ is the currently vacant position. The current solution is stored in array $x$. For $i=1,2,\cdots,nm+n+m$, the move of step $i$ is stored in $x[i-1]$. This means that we move the checker located at positions $x[i-1]$ to the current vacant position and leaving the positions $x[i-1]$ the new vacant position. A recursive function call $\CALL{Backtrack}{1,n+1}$ will generate all optimal solutions which move checkers from initial state to a final state in $nm+n+m$ steps.

It is not difficult to generate all optimal solutions of the game with small size by the backtracking algorithm described above.

\begin{pseudocode}[shadowbox]{Backtrack}{i,e}
 \COMMENT{ Generate all optimal solutions }\\
 \IF i>nm+n+m \THEN
    \BEGIN
       \IF e=m+1  \AND \text{final state reached }
          \THEN \text{output current solution}\\
    \END
 \ELSE
   \BEGIN
       \IF e>2 \AND jump(b,w,r)  \text{ feasible} \THEN
         \BEGIN
           x[i-1]\GETS e-2\\
           \text{move checker at position $e-2$ to vacant}\\
           \CALL{Backtrack}{i+1,e-2}\\
           \text{move checker at position $e$ to vacant}\\
         \END\\

       \IF e<n+m \AND jump(b,w,l)  \text{ feasible} \THEN
         \BEGIN
           x[i-1]\GETS e+2\\
           \text{move checker at position $e+2$ to vacant}\\
           \CALL{Backtrack}{i+1,e+2}\\
           \text{move checker at position $e$ to vacant}\\
         \END\\

       \IF e>1 \AND slide(b,r) \text{ feasible} \THEN
         \BEGIN
           x[i-1]\GETS e-1\\
           \text{move checker at position $e-1$ to vacant}\\
           \CALL{Backtrack}{i+1,e-1}\\
           \text{move checker at position $e$ to vacant}\\
         \END\\

       \IF e<n+m+1 \AND slide(w,l)  \text{ feasible} \THEN
         \BEGIN
           x[i-1]\GETS e+1\\
           \text{move checker at position $e+1$ to vacant}\\
           \CALL{Backtrack}{i+1,e+1}\\
           \text{move checker at position $e$ to vacant}\\
         \END\\
   \END
\end{pseudocode}

\section{A Linear Time Construction Algorithm}

The backtracking algorithm described in the previous section can produce all optimal solutions for the game with fixed size. It generally works only for small size. In this section, we will present a linear time construction algorithm which can produce all optimal solutions in linear time for very large size $n+m$.
The Decrease-and-Conquer strategy [4] for algorithm design is exploited to design our new algorithm.

Without loss of generality, we assume $n\geq m$ in the following discussion. Since there are only 4 possible moves $slide(w,l)$, $slide(b,r)$, $jump(b,w,l)$, and $jump(b,w,r)$, we can simplify our notation for these 4 moves to $slide(l)$, $slide(r)$, $jump(l)$, and $jump(r)$ in the following discussion.

\subsection{A special case of the problem}

We first focus on the special case of $n=m$. If we denote a black checker by $b$, a white checker by $w$, and the vacant position by $O$, then any status of the checker board can be specified by a sequence consisting of characters $b, w$ and $O$.
The special case of our problem is then equivalent to transforming the initial sequence ${\overbrace{b\cdots b}^m} O {\overbrace{w\cdots w}^m}$ to the sequence ${\overbrace{w\cdots w}^m} O {\overbrace{b\cdots b}^m}$ in the minimum number of steps.

We have noticed that a key status of the checker board can be reached from the initial status with minimum number of steps.

\begin{lemma}\label{le2}
The initial status of the checker board ${\overbrace{b\cdots b}^m} O {\overbrace{w\cdots w}^m}$ can be transformed to one of the status of the checker board $O \overbrace{bw\cdots bw}^{2m}$ or $\overbrace{bw\cdots bw}^{2m} O$ in $\frac{m(m+1)}{2}$ steps.
\end{lemma}
\noindent{\bf Proof.}
We can design a recursive algorithm to solve this problem as follows.

\begin{pseudocode}[shadowbox]{move1}{t}
 \IF t>1 \THEN \CALL{move1}{t-1}\\
     \COMMENT{ $t-1$ jumps }\\
     \FOR i \GETS 1 \TO t-1 \DO
       \CALL{jump}{dir}\\
     \COMMENT{ 1 slide}\\
     \CALL{slide}{dir}\\
     \COMMENT{ change moving direction }\\
     \CALL{change}{dir}\\
\end{pseudocode}

In the algorithm described above, the parameter $t$ is the recursion depth, or the number of black checkers to be treated. The variable $dir$ is used to determine the current moving direction. Its value $l$ indicates the checker should be moved left, otherwise the checker should be moved right. The initial value of $dir$ can be set to $l$ or $r$, which will lead to different moving sequences. The current direction $dir$ can be changed in $O(1)$ time by the algorithm \CALL{change}{$dir$} as follows.

\begin{pseudocode}[shadowbox]{change}{dir}
     \IF dir=r \THEN dir \GETS l
       \ELSE dir \GETS r\\
\end{pseudocode}

Based on the algorithm above, the Lemma can be proved by induction.
The moving steps generated by the algorithm \CALL{move1}{} for the first two easy cases of $m=1$ and $m=2$ are shown in Table 2 and Table 3. The number of steps for these two cases are 1 and 3 respectively. The lemma is correct for the base cases.

Assume that, the Lemma is true for $m<t$. For the case of $m=t$, the algorithm \CALL{move1}{$t-1$} is applied first and the status of the checkerboard is transformed to $bO \overbrace{bw\cdots bw}^{2t-2}w$ or $b\overbrace{bw\cdots bw}^{2t-2} Ow$ depending on the initial value of $dir$. Then, $t-1$ jumps followed by 1 slide of the algorithm \CALL{move1}{} will transform the status of the checkerboard to $\overbrace{bw\cdots bw}^{2t} O$ or $O \overbrace{bw\cdots bw}^{2t}$. The algorithm \CALL{move1}{$t-1$} needs $(t-1)t/2$ steps by the induction hypothesis, so the number of steps used by the algorithm \CALL{move1}{$t$} is $$(t-1)t/2+t-1+1=(t-1)t/2+t=t(t+1)/2$$

The proof is completed. $\Box$

The key status of the checkerboard $\overbrace{bw\cdots bw}^{2m} O$ or $O \overbrace{bw\cdots bw}^{2m}$ can be transformed to
another key status of the checkerboard $O \overbrace{wb\cdots wb}^{2m}$ or $\overbrace{wb\cdots wb}^{2m} O$ readily by $m$ jumps. Any of these two status of the checkerboard can then be transformed to the final status ${\overbrace{w\cdots w}^m} O {\overbrace{b\cdots b}^m}$. This problem is exactly the inverse problem of Lemma 2.

\begin{lemma}\label{le3}
The key status of the checkerboard $O \overbrace{wb\cdots wb}^{2m}$ or $\overbrace{wb\cdots wb}^{2m} O$
can be transformed to the final status ${\overbrace{w\cdots w}^m} O {\overbrace{b\cdots b}^m}$ in $\frac{m(m+1)}{2}$ steps.
\end{lemma}
\noindent{\bf Proof.}

We can design a recursive algorithm to solve this problem, which is exactly a reversed algorithm of the algorithm \CALL{move1}{}.

\begin{pseudocode}[shadowbox]{move4}{t}
   \COMMENT{ change moving direction }\\
   \CALL{change}{dir}\\
   \COMMENT{ 1 slide}\\
   \CALL{slide}{dir}\\
   \COMMENT{ $t-1$ jumps }\\
   \FOR i \GETS 1 \TO t-1 \DO
       \CALL{jump}{dir}\\
   \COMMENT{ recursive call }\\
   \IF t>1 \THEN \CALL{move4}{t-1}\\
\end{pseudocode}

In the algorithm described above, the parameter $t$ is the recursion depth, or the number of black checkers to be treated. The variable $dir$ is used to determine the current moving direction. Its initial value is retained from previous computation.

Based on the algorithm above, the Lemma can be proved by induction.
The first two easy cases of $m=1$ and $m=2$ are similar to the cases of Table 2 and Table 3. The number of steps for these two cases are 1 and 3 respectively. The lemma is correct for the base cases.

Assume that, the Lemma is true for $m<t$. In the case of $m=t$, the algorithm \CALL{move4}{$t$} implement 1 slide and $t-1$ jumps first. The 2 key status of the checkerboard $O \overbrace{wb\cdots wb}^{2m}$ or $\overbrace{wb\cdots wb}^{2m} O$ can then be transformed to the status of the checkerboard $w\overbrace{wb\cdots wb}^{2m-2} Ob$ or $wO \overbrace{wb\cdots wb}^{2m-2}b$ respectively. Then a recursive call \CALL{move4}{$t-1$} is applied to transform the checkerboard to the final status ${\overbrace{w\cdots w}^m} O {\overbrace{b\cdots b}^m}$.

The algorithm \CALL{move4}{$t-1$} needs $(t-1)t/2$ steps by the induction hypothesis, so the number of steps used by the algorithm \CALL{move4}{$t$} is $$1+t-1+(t-1)t/2=t+(t-1)t/2=t(t+1)/2$$

The proof is completed. $\Box$

\begin{table}
\caption{Move1 for the easy case of $m=1$}
\begin{center}
\begin{tabular}{|c|c|c|c|}
\hline
\textit{Direction} & \textit{Step} & \textit{Move} & \textit{Status}\\\hline
\multirow{3}*{$dir=1$}
& 0 & &\raisebox{-.6ex}{\FilledCircle \ \Square \ \Circle}\\ \cline{2-4}
& 1 & $slide(l)$ &\raisebox{-.6ex}{\FilledCircle \ \Circle \ \Square}\\\hline
\multirow{3}*{$dir=-1$}
& 0 & &\raisebox{-.6ex}{\FilledCircle \ \Square \ \Circle}\\ \cline{2-4}
& 1 & $slide(r)$ &\raisebox{-.6ex}{\Square \ \FilledCircle \ \Circle}\\\hline
\end{tabular}
\end{center}
\end{table}

\begin{table}
\caption{Move1 for the easy case of $m=2$}
\begin{center}
\begin{tabular}{|c|c|c|c|}
\hline
\textit{Direction} & \textit{Step} & \textit{Move} & \textit{Status}\\\hline
\multirow{6}*{$dir=1$}
& 0 & &\raisebox{-.6ex}{\FilledCircle \ \FilledCircle \ \Square \ \Circle\ \Circle}\\ \cline{2-4}
& 1 & $slide(l)$ &\raisebox{-.6ex}{\FilledCircle \ \FilledCircle \ \Circle \ \Square\ \Circle}\\\cline{2-4}
& 2 & $jump(r)$ &\raisebox{-.6ex}{\FilledCircle\ \Square \ \Circle \ \FilledCircle\ \Circle}\\ \cline{2-4}
& 3 & $slide(r)$ &\raisebox{-.6ex}{\Square \ \FilledCircle\ \Circle \ \FilledCircle\ \Circle}\\ \hline
\multirow{6}*{$dir=-1$}
& 0 & &\raisebox{-.6ex}{\FilledCircle \ \FilledCircle \ \Square \ \Circle\ \Circle}\\ \cline{2-4}
& 1 & $slide(r)$ &\raisebox{-.6ex}{\FilledCircle \  \Square\ \FilledCircle \ \Circle\ \Circle}\\\cline{2-4}
& 2 & $jump(l)$ &\raisebox{-.6ex}{\FilledCircle \  \Circle\ \FilledCircle \ \Square\ \Circle}\\ \cline{2-4}
& 3 & $slide(l)$ &\raisebox{-.6ex}{\FilledCircle \  \Circle\ \FilledCircle \ \Circle\ \Square}\\ \hline
\end{tabular}
\end{center}
\end{table}

The 3 stages of the algorithms can now be combined into a new algorithm to solve our problem for the special case of $n=m$ as follows.

\begin{pseudocode}[shadowbox]{move}{m,d}
   \COMMENT{ initial moving direction }\\
   dir \GETS d\\
   \COMMENT{ first stage }\\
   \CALL{move1}{m}\\
   \COMMENT{ $m$ jumps }\\
   \FOR i \GETS 1 \TO m \DO
       \CALL{jump}{dir}\\
   \COMMENT{ last stage }\\
   \CALL{move4}{m}\\
\end{pseudocode}

The algorithm requires $m(m+1)/2+m+m(m+1)/2=m^2+2m$ steps. It has been known that $m^2+2m$ is a lower bound to solve the game consisting of $m$ black checkers and $m$ white checkers by Theorem~\ref{th1}.
Therefore, our algorithm is optimal to solve the game for the special case of $n=m$. From this point, we can also claim that the algorithms \CALL{move1}{} and \CALL{move4}{} presented in the proofs of Lemma 2 and Lemma 3 are also optimal. Otherwise, there must be an algorithm to solve the problem in less than $m^2+2m$ steps and this is impossible.

\subsection{The algorithm for the general case of the problem}

We have discussed the special case of $n=m$. In this subsection, we will discuss the general cases $n>m$ of the problem.
In these general cases, $n-m>0$.

We can first use the algorithm  \CALL{move1}{} to transform the checkerboard to the status $\overbrace{b\cdots b}^{n-m} O \overbrace{bw\cdots bw}^{2m}$ or $\overbrace{b\cdots b}^{n-m} \overbrace{bw\cdots bw}^{2m} O$ in $\frac{m(m+1)}{2}$ steps.
Then $m$ jumps are applied to transform the checkerboard to the status $\overbrace{b\cdots b}^{n-m} O \overbrace{wb\cdots wb}^{2m}$ or $\overbrace{b\cdots b}^{n-m} \overbrace{wb\cdots wb}^{2m} O$.

At this point, we have to try to move the leftmost $n-m$ black checkers to the rightmost $n-m$ positions. It is not difficult to do this by a simple algorithm similar to the algorithm \CALL{move1}{}.

\begin{lemma}\label{le4}
The key status of the checkerboard $\overbrace{b\cdots b}^{n-m} O \overbrace{wb\cdots wb}^{2m}$ or $\overbrace{b\cdots b}^{n-m} \overbrace{wb\cdots wb}^{2m} O$ can be transformed to the status $\overbrace{wb\cdots wb}^{2m} O\overbrace{b\cdots b}^{n-m}$ or $O \overbrace{wb\cdots wb}^{2m}\overbrace{b\cdots b}^{n-m}$ in $(n-m)(m+1)$ steps.
\end{lemma}
\noindent{\bf Proof.}

We can design a recursive algorithm to solve this problem as follows.

\begin{pseudocode}[shadowbox]{move3}{t}
   \COMMENT{ 1 slide to right}\\
   \CALL{slide}{r}\\
   \COMMENT{ change jumping direction }\\
   \CALL{change}{dir}\\
   \COMMENT{ $m$ jumps }\\
   \FOR i \GETS 1 \TO m \DO
       \CALL{jump}{dir}\\
   \COMMENT{ recursive call }\\
   \IF t>1 \THEN \CALL{move3}{t-1}\\
\end{pseudocode}

In the algorithm described above, the parameter $t$ is the recursion depth, or the number of black checkers to be moved to the right most positions. The variable $dir$ is used to determine the current jumping direction. Its initial value is retained from previous computation.

Based on the algorithm above, the Lemma can be proved by induction on $n-m$.
When $n-m=1$, we have to move the leftmost black checker to rightmost. We first make a slide right, then $m$ jumps followed as described by the algorithm \CALL{move3}{}. The status of the checkerboard will be changed to $\overbrace{wb\cdots wb}^{2m} Ob$ or $O \overbrace{wb\cdots wb}^{2m}b$. It costs $m+1$ steps. The lemma is correct for the base case of $n-m=1$.

Assume that, the Lemma is true for $n-m<t$. For the case of $n-m=t$, the algorithm \CALL{move3}{$t$} implement 1 slide and $m$ jumps first. The 2 key status of the checkerboard $\overbrace{b\cdots b}^{t} O \overbrace{wb\cdots wb}^{2m}$ or $\overbrace{b\cdots b}^{t} \overbrace{wb\cdots wb}^{2m} O$ can then be transformed to the status of the checkerboard $\overbrace{b\cdots b}^{t-1} \overbrace{wb\cdots wb}^{2m} Ob$ or $\overbrace{b\cdots b}^{t-1} O \overbrace{wb\cdots wb}^{2m}b$ respectively. Then a recursive call \CALL{move3}{$t-1$} is applied to transform the checkerboard to the status $\overbrace{wb\cdots wb}^{2m} O\overbrace{b\cdots b}^{t}$ or $O \overbrace{wb\cdots wb}^{2m}\overbrace{b\cdots b}^{t}$.

The algorithm \CALL{move3}{$t-1$} needs $(t-1)(m+1)$ steps by the induction hypothesis, so the number of steps used by the algorithm \CALL{move3}{$t$} is $$m+1+(t-1)(m+1)=t(m+1)$$

The proof is completed. $\Box$

The 4 stages of the algorithms can now be combined into a new algorithm to solve our problem for the general cases of $n\geq m$ as follows.

\begin{pseudocode}[shadowbox]{move}{n,m,d}
   \COMMENT{ initial moving direction }\\
   dir \GETS d\\
   \COMMENT{ first stage }\\
   \CALL{move1}{m}\\
   \COMMENT{ second stage }\\
   \FOR i \GETS 1 \TO m \DO
       \CALL{jump}{dir}\\
   \COMMENT{ third stage }\\
   \IF n-m>0 \THEN \CALL{move3}{n-m}\\
   \COMMENT{ last stage }\\
   \CALL{move4}{m}\\
\end{pseudocode}

By Lemma 2, Lemma 3 and Lemma 4 we know that the algorithm requires $m(m+1)/2+m+(n-m)(m+1)+m(m+1)/2=nm+n+m$ steps. It has been known from Theorem 1 that $nm+n+m$ is a lower bound to solve the game consisting of $n$ black checkers and $m$ white checkers.
Therefore, our algorithm is optimal to solve the game for the general cases of $n\geq m$. We can also claim that the algorithm \CALL{move3}{} is also optimal. Otherwise, there must be an algorithm to solve the problem in less than $nm+n+m$ steps and this is impossible.

\begin{theo}\label{th2}
The algorithm \CALL{move}{n,m,d} requires $nm+n+m$ steps to solve the general moving checkers game consisting of $n$ black checkers and $m$ white checkers, and the algorithm is optimal.
\end{theo}

\subsection{Remove recursions }

The algorithms \CALL{move1}{}, \CALL{move3}{} and \CALL{move4}{} are all recursive algorithms. The recursions of these algorithms can be easily removed by only one \FOR loop.

The equivalent iterative algorithm for solving the general moving checkers game consisting of $n$ black checkers and $m$ white checkers can be described as follows.

\begin{pseudocode}[shadowbox]{iterative\_move}{n,m,d}
   \COMMENT{ initial moving direction }\\
   dir \GETS d\\
   \COMMENT{ stage 1}\\
   \FOR i \GETS 1 \TO m \DO
     \BEGIN
       \FOR j \GETS 1 \TO i-1 \DO
          \CALL{jump}{dir}\\
       \CALL{slide}{dir}\\
       \CALL{change}{dir}\\
     \END\\
   \COMMENT{ stage 2 }\\
   \FOR i \GETS 1 \TO m \DO
       \CALL{jump}{dir}\\
   \COMMENT{ stage 3 }\\

   \FOR i \GETS 1 \TO n-m \DO
     \BEGIN
       \CALL{slide}{r}\\
       \CALL{change}{dir}\\
       \FOR j \GETS 1 \TO m \DO
          \CALL{jump}{dir}\\
     \END\\
   \COMMENT{ stage 4 }\\

   \FOR i \GETS m \DOWNTO 1 \DO
     \BEGIN
       \CALL{change}{dir}\\
       \CALL{slide}{dir}\\
       \FOR j \GETS 1 \TO i-1 \DO
          \CALL{jump}{dir}\\
     \END\\
\end{pseudocode}

\section{The Explicit Solutions to the Problem }

The optimal solution found by the algorithm \CALL{move}{} or \CALL{iterative\_move}{} can be presented by a vector $x$. For $i=1,2,\cdots,nm+n+m$, the step $i$ of the optimal move sequence is given by $x_i$. This means that the checker located at position $x_i$ will be moved in step $i$ to the current vacant positions and leaving the positions $x_i$ the new vacant positions. This can also be viewed that $x$ is a function of $i$, which is called a move function.
In this section we will discuss the explicit expression of function $x$.

If we denote $x_0=n+1$ and

\beql{eq1}
d_i=x_{i-1}-x_i, 1\leq i\leq nm+n+m
\eeq

then the vector $d$ will be a move direction function of the corresponding move sequence.

A related function $t$ can then be defined as $t_i=\sum_{j=1}^i d_j, 1\leq i\leq nm+n+m$.

Since $$t_i=\sum_{j=1}^i d_j=\sum_{j=1}^i(x_{j-1}-x_j)=x_0-x_i=n+1-x_i$$

we have

\beql{eq2}
x_i=n+1-t_i, 1\leq i\leq nm+n+m
\eeq

Therefore, our task is equivalent to compute the function $t$ efficiently.

In this section, the functions $x$ and $t$ will be divided into three parts. The first part is corresponding to the first two stages of the algorithm \CALL{iterative\_move}{} presented in the last section. The second part is corresponding to the stage 3 of the algorithm \CALL{iterative\_move}{} and the third part is corresponding to the stage 4.

\subsection{The first part of the solution}

Similar to the initial value of $dir$ which can be set to $l$ or $r$, the first move direction $d_1$ can be set to 1 or -1.
If we set $d_1=1$, then from the algorithm \CALL{iterative\_move}{} presented in the last section, the move direction sequence for the stage 1 and 2 must be $1,-2,-1,2,2,1,-2,-2,-2,\cdots, (-1)^{m-1}, \overbrace{2(-1)^m,\cdots,2(-1)^m}^{m}$. This move direction sequence can be divided into $m$ sections as $$\overbrace{1,-2}^{2},\overbrace{-1,2,2}^{3},\overbrace{1,-2,-2,-2}^{4},\cdots,  \overbrace{(-1)^{m-1},2(-1)^m,\cdots,2(-1)^m}^{m+1}$$

The section $j$ consists of 1 slide and $j$ jumps and thus has a size of $j+1$.

The total length of the sequence is therefore $s_1=\sum_{j=1}^m (j+1)=m(m+3)/2$.
Our task is now to find $t_i=\sum_{j=1}^i d_j$ quickly for each $1\leq i\leq s_1$.

If we denote the $j+1$ elements of the section $j$ as $a_{tj}, 1\leq t\leq j+1$, and the sum of section $j$ as $a_j=\sum_{t=1}^{j+1}a_{tj}, j=1,\cdots,m$, then it is not difficult to see that for each $j=1,\cdots,m$,

\beql{eq3}
a_{tj}=\left\{\begin{array}{ll}
(-1)^{j-1} & t=1\\
2(-1)^j & t>1\\
\end{array}\right.
\eeq

and for $1\leq k\leq j+1$,

\beql{eq4}
\sum_{t=1}^ka_{tj}=(-1)^j(2k-3)
\eeq

Therefore, $a_j=(-1)^j(2j-1), j=1,\cdots,m$. These $m$ sums form an alternating sequence
$$-1, 3, -5, \cdots, (-1)^m(2m-1)$$

For each $1\leq k\leq m$, we have,

\beql{eq5}
\sum_{j=1}^ka_j=\sum_{j=1}^k(-1)^j(2j-1)=(-1)^kk
\eeq

The steps in each section must be
$$\overbrace{1,2}^{1},\overbrace{3,4,5}^{2},\overbrace{6,7,8,9}^{3},\cdots,  \overbrace{(m-1)(m+2)/2+1,\cdots,m(m+3)/2}^{m}$$

If we denote the $j+1$ steps of the section $j$ as $b_{tj}, 1\leq t\leq j+1$, and the boundary of section $j$ as $b_j=b_{(j+1)j}, j=1,\cdots,m$, then it is not difficult to see that for each $j=1,\cdots,m$,

\beql{eq6}
\left\{\begin{array}{ll}
b_j=j(j+3)/2 & 1\leq j\leq m\\
b_{tj}=b_{j-1}+t & 1\leq t\leq j+1, 1\leq j\leq m\\
\end{array}\right.
\eeq

For any integer $1\leq i\leq b_m$, the corresponding integer $k$ such that the integer $i$ falls into the section $k$ can be computed by a function $f_1(x)$ as follows.

\begin{lemma}\label{le5}
Let $f_1(x)=\frac{\sqrt{8x+1}-1}{2}$. For any integer $1\leq i\leq b_m$, it must be a step number in the section $k=\lfloor f_1(i)\rfloor $.
\end{lemma}
\noindent{\bf Proof.}

It can be seen readily that function $f_1(x)$ is a strictly increasing function on $(0,+\infty)$. For each section $k, 1\leq k\leq m$, its first step number is $b_{k-1}+1=(k-1)(k+2)/2+1$ and it satisfies

$$f_1((k-1)(k+2)/2+1)=\frac{\sqrt{4(k-1)(k+2)+9}-1}{2}=\frac{\sqrt{(2k+1)^2}-1}{2}=k$$

Therefore, for each integer $i$ in the section $k$, we have, $k\leq f_1(i)< k+1$. This means $\lfloor f_1(i)\rfloor=k$.

The proof is completed. $\Box$

From Lemma~\ref{le5} and formula (4) and (5), we can now compute $t_i=\sum_{j=1}^i d_j, 1\leq i\leq s_1$ as follows.

Let $\alpha=\lfloor f_1(i)\rfloor$, then,
$t_i=\sum_{j=1}^i d_j=\sum_{j=1}^{\alpha-1}a_j+\sum_{j=b_{\alpha-1}+1}^id_j=(-1)^{\alpha-1}(\alpha-1)+(-1)^{\alpha}(2(i-b_{\alpha-1})-3)$.
It follows that for each $1\leq i\leq s_1$,

\beql{eq7}
t_i=(-1)^\alpha(2i-\alpha(\alpha+2))
\eeq

where, $\alpha=\lfloor \frac{\sqrt{8i+1}-1}{2}\rfloor$.

It follows from formula (2) that for each $1\leq i\leq s_1$,

\beql{eq8}
x_i=n+1-(-1)^\alpha(2i-\alpha(\alpha+2))
\eeq

If we set $d_1=-1$, a similar result will be obtained. In this case, we have,
\beql{eq9}
x_i=n+1+(-1)^\alpha(2i-\alpha(\alpha+2))
\eeq

Combine these two cases, we conclude that,
\beql{eq10}
x_i=n+1-d_1(-1)^\alpha(2i-\alpha(\alpha+2))
\eeq

\subsection{The second part of the solution}
If we set $d_1=1$, then according to the algorithm \CALL{iterative\_move}{} presented in the last section, the move direction sequence for the stage 3 must be

$1,\overbrace{2(-1)^{m+1},\cdots,2(-1)^{m+1}}^{m},1,\overbrace{2(-1)^{m+2},\cdots,2(-1)^{m+2}}^{m},\cdots, 1,\overbrace{2(-1)^{n},\cdots,2(-1)^{n}}^{m}$.

This move direction sequence can be divided naturally into $n-m$ sections.
The section $j$ consists of 1 slide and $m$ jumps and thus has a size of $m+1$.
The total length of the sequence is therefore $s_2=(n-m)(m+1)$.
Our task for this part is now to find $t_i=\sum_{j=1}^i d_j$ quickly for each $s_1+1\leq i\leq s_1+s_2$.

If we denote the $m+1$ elements of the section $j$ as $a_{tj}, 1\leq t\leq m+1$, and the sum of section $j$ as $a_j=\sum_{t=1}^{m+1}a_{tj}, j=1,\cdots,n-m$, then it is not difficult to see that for each $j=1,\cdots,n-m$,

\beql{eq11}
a_{tj}=\left\{\begin{array}{ll}
1 & t=1\\
2(-1)^{m+j} & t>1\\
\end{array}\right.
\eeq

and for $1\leq k\leq m+1$,

\beql{eq12}
\sum_{t=1}^ka_{tj}=1+(-1)^{m+j}(2k-2)
\eeq

Therefore, $a_j=1+2m(-1)^{m+j}, j=1,\cdots,n-m$. These $n-m$ sums form an alternating sequence
$$1+2m(-1)^{m+1}, 1+2m(-1)^{m+2}, \cdots, 1+2m(-1)^{n}$$

For each $1\leq k\leq m$, we have,

\beql{eq13}
\sum_{j=1}^ka_j=k+\sum_{j=1}^k2m(-1)^{m+j}=k+m(-1)^{m+k}-m(-1)^{m}
\eeq

If we set $j=i-s_1$, then the steps in each section must be
$$\overbrace{1,\cdots, m+1}^{m+1},\overbrace{m+2,\cdots,2m+2}^{m+1},\cdots,  \overbrace{(n-m-1)(m+1)+1,\cdots,(n-m)(m+1)}^{m+1}$$

If we denote the $m+1$ steps of the section $j$ as $b_{tj}, 1\leq t\leq m+1$, and the boundary of section $j$ as $b_j=b_{(m+1)j}, j=1,\cdots,n-m$, then it is not difficult to see that for each $j=1,\cdots,n-m$,

\beql{eq14}
\left\{\begin{array}{ll}
b_j=j(m+1) & 1\leq j\leq n-m\\
b_{tj}=b_{j-1}+t & 1\leq t\leq m+1, 1\leq j\leq n-m\\
\end{array}\right.
\eeq

For any integer $1\leq j\leq b_m$, the corresponding integer $k$ such that the integer $j$ falls into the section $k$ can be computed by a function $f_2(x)$ as follows.

\begin{lemma}\label{le6}
Let $f_2(x)=\frac{x+m}{m+1}$. For any integer $1\leq j\leq b_m$, it must be a step number in the section $k=\lfloor f_2(j)\rfloor $.
\end{lemma}
\noindent{\bf Proof.}

It can be seen readily that function $f_2(x)$ is a strictly increasing function on $(0,+\infty)$. For each section $k, 1\leq k\leq n-m$, its first step number is $b_{k-1}+1=(k-1)(m+1)+1$ and it satisfies

$$f_2((k-1)(m+1)+1)=\frac{(k-1)(m+1)+1+m}{m+1}=\frac{k(m+1)}{m+1}=k$$

Therefore, for each integer $j$ in the section $k$, we have, $k\leq f_2(j)< k+1$. This means $\lfloor f_2(j)\rfloor=k$.

The proof is completed. $\Box$

From Lemma~\ref{le5} and formula (13) and (14), we can now compute $t_i=\sum_{j=1}^i d_j, s_1+1\leq i\leq s_1+s_2$ as follows.

Let $r=i-s_1$, $\beta=\lfloor f_2(r)\rfloor$,  and  $p=r-(\beta-1)(m+1)$ then,
$t_i=\sum_{j=1}^i d_j=t_{s_1}+\sum_{j=1}^{\beta-1}a_j+\sum_{j=1}^{p}a_{j\beta}$.

Therefore

$$\begin{array}{ll}
t_i-t_{s_1}&=\beta-1+m(-1)^{m+\beta-1}-m(-1)^{m}+1+(-1)^{m+\beta}(2p-2)\\
&=\beta+(-1)^{m+\beta}(2p-2)-m((-1)^{m+\beta}+(-1)^{m+2\beta})\\
&=\beta+(-1)^{m+\beta}(2p-2-m(1+(-1)^{\beta}))
\end{array}$$

It follows that for each $s_1+1\leq i\leq s_1+s_2$,

\beql{eq15}
t_i=t_{s_1}+\beta+(-1)^{m+\beta}(2p-2-m(1+(-1)^{\beta}))
\eeq

where, $\beta=\lfloor \frac{i-s_1+m}{m+1} \rfloor$, and $p=i-s_1-(\beta-1)(m+1)$.

It follows from formula (2) that for each $s_1+1\leq i\leq s_1+s_2$,

\beql{eq16}
x_i=n+1-t_{s_1}-\beta-(-1)^{m+\beta}(2p-2-m(1+(-1)^{\beta}))
\eeq

If we set $d_1=-1$, a similar result will be obtained. In this case, we have,
\beql{eq17}
x_i=n+1-t_{s_1}-\beta+(-1)^{m+\beta}(2p-2-m(1+(-1)^{\beta}))
\eeq

Combine these two cases, we conclude that,
\beql{eq18}
x_i=x_{s_1}-\beta-d_1(-1)^{m+\beta}(2p-2-m(1+(-1)^{\beta}))
\eeq

\subsection{The third part of the solution}
According to the algorithm \CALL{iterative\_move}{} presented in the last section, if $d_1=1$, then the move direction sequence for the stage 4 must be

$(-1)^{n+1}(1,\overbrace{2,\cdots,2}^{m-1},-1,\overbrace{-2,\cdots,-2}^{m-2},\cdots, (-1)^{m-1})$.

This move direction sequence can be divided naturally into $m$ sections.
The section $j$ consists of 1 slide and $m-j$ jumps and thus has a size of $m-j+1$.
The total length of the sequence is therefore $s_3=m(m+1)/2$.
Our task for this part is now to find $t_i=\sum_{j=1}^i d_j$ quickly for each $s_1+s_2+1\leq i\leq s_1+s_2+s_3=nm+n+m$.

If we denote the $m-j+1$ elements of the section $j$ as $a_{tj}, 1\leq t\leq m-j+1$, and the sum of section $j$ as $a_j=\sum_{t=1}^{m-j+1}a_{tj}, j=1,\cdots,m$, then it is not difficulty to see that for each $j=1,\cdots,m$,

\beql{eq19}
a_{tj}=\left\{\begin{array}{ll}
(-1)^{n+j} & t=1\\
2(-1)^{n+j} & t>1\\
\end{array}\right.
\eeq

and for $1\leq k\leq m-j+1$,

\beql{eq20}
\sum_{t=1}^ka_{tj}=(-1)^{n+j}(2k-1)
\eeq

Therefore, $a_j=(-1)^{n+j}(2(m-j)+1), j=1,\cdots,m$. These $m$ sums form an alternating sequence
$$(-1)^{n+1}((2m-1), -(2m-3), \cdots, (-1)^{m-1})$$

For each $1\leq k\leq m$, we have,

$$
\sum_{j=1}^ka_j=\sum_{j=1}^k(-1)^{n+j}(2m-(2j-1))=(-1)^n(2m((-1)^k-1)/2)-(-1)^kk)
$$

Therefore,

\beql{eq21}
\sum_{j=1}^ka_j=(-1)^n((-1)^k(m-k)-m)
\eeq

If we set $j=i-s_1-s_2$, then the step numbers in each sections must be
$$\overbrace{1,\cdots, m}^{m},\overbrace{m+1,\cdots,2m-1}^{m-1},\cdots,  \overbrace{m(m+1)/2}^{1}$$

If we denote the $m-j+1$ steps of the section $j$ as $b_{tj}, 1\leq t\leq m-j+1$, and the boundary of section $j$ as $b_j=b_{(m-j+1)j}, j=1,\cdots,m$, then it is not difficulty to see that for each $j=1,\cdots,m$,

\beql{eq22}
\left\{\begin{array}{ll}
b_j=j(m+1)-j(j+1)/2 & 1\leq j\leq m\\
b_{tj}=b_{j-1}+t & 1\leq t\leq m-j+1, 1\leq j\leq m\\
\end{array}\right.
\eeq

For any integer $1\leq j\leq b_m$, the corresponding integer $k$ such that the integer $j$ falls into the section $k$ can be computed by a function $f_3(x)$ as follows.

\begin{lemma}\label{le7}
Let $f_3(x)=m-\sqrt{m(m+1)-2x+9/4}+3/2$. For any integer $1\leq j\leq b_m$, it must be a step number in the section $k=\lfloor f_3(j)\rfloor $.
\end{lemma}
\noindent{\bf Proof.}

It can be seen readily that function $f_2(x)$ is a strictly increasing function on $(0,m(m+1)/2]$. For each section $k, 1\leq k\leq m$, its first step number is $b_{k-1}+1=(k-1)(m+1)-k(k-1)/2+1$ and it satisfies

$$\begin{array}{ll}
f_3((k-1)(m+1)-k(k-1)/2+1)& \\
=m+3/2-\sqrt{m(m+1)-2(k-1)(m+1)+k(k-1)-2+9/4} & \\
=m+3/2-\sqrt{(k-m-3/2)^2}=k &
\end{array}$$

Therefore, for each integer $j$ in the section $k$, we have, $k\leq f_3(j)< k+1$. This means $\lfloor f_3(j)\rfloor=k$.

The proof is completed. $\Box$

From Lemma~\ref{le7} and formula (21) and (22), we can now compute $t_i=\sum_{j=1}^i d_j, s_1+s_2+1\leq i\leq nm+n+m$ as follows.

Let $r=i-s_1-s_2$, $\gamma=\lfloor f_3(r)\rfloor$,  and  $q=r-(\gamma-1)(m+1)+\gamma(\gamma-1)/2$ then,

$t_i=\sum_{j=1}^i d_j=t_{s_2}+\sum_{j=1}^{\gamma-1}a_j+\sum_{j=1}^{q}a_{j\gamma}$.

Therefore

$$\begin{array}{ll}
t_i-t_{s_2}&=(-1)^n((-1)^{\gamma-1}(m-\gamma+1)-m)+(-1)^{n+\gamma}(2q-1)\\
&=(-1)^{n+\gamma}(\gamma+2q-m-2)-m(-1)^n
\end{array}$$

It follows that for each $s_1+s_2+1\leq i\leq nm+n+m$,

\beql{eq23}
t_i=t_{s_2}+(-1)^{n+\gamma}(\gamma+2q-m-2)-m(-1)^n
\eeq

where, $\gamma=\lfloor m-\sqrt{m(m+1)-2(i-s_1-s_2)+9/4}+3/2 \rfloor$, and $q=i-s_1-s_2-(\gamma-1)(m+1)+\gamma(\gamma-1)/2$.

It follows from formula (2) that for each $s_1+s_2+1\leq i\leq nm+n+m$,

\beql{eq24}
x_i=n+1-t_{s_2}-(-1)^{n+\gamma}(\gamma+2q-m-2)+m(-1)^n
\eeq

If we set $d_1=-1$, a similar result will be obtained. In this case, we have,
\beql{eq25}
x_i=n+1-t_{s_2}+(-1)^{n+\gamma}(\gamma+2q-m-2)-m(-1)^n
\eeq

Combine these two cases, we conclude that,
\beql{eq26}
x_i=x_{s_2}-d_1((-1)^{n+\gamma}(\gamma+2q-m-2)-m(-1)^n)
\eeq

Summing up, the explicit optimal solutions for solving the general game of shifting the checkers consisting of $n$ black checkers and $m$ white checkers can be given in three parts as shown in the following Theorem.

\begin{theo}\label{th3}
In the general game of shifting the checkers consisting of $n$ black checkers and $m$ white checkers, its optimal move steps  $x_i,1\leq i\leq nm+n+m$, can be expressed explicitly as follows.

\beql{eq27}
x_i=\left\{\begin{array}{ll}
n+1-(-1)^{\alpha}d(2i-\alpha(\alpha+2)) & 1\leq i\leq s_1\\
x_{s_1}-\beta-(-1)^{m+\beta}d(2p-2-m(1+(-1)^{\beta})) & s_1+1\leq i\leq s_1+s_2\\
x_{s_2}-d((-1)^{n+\gamma}(\gamma+2q-m-2)-m(-1)^n) & s_1+s_2+1\leq i\leq nm+n+m\\
\end{array}\right.
\eeq

where, $d$ is the first move direction, and

\beql{eq28}
\left\{\begin{array}{l}
s_1=m(m+3)/2\\
s_2=(n-m)(m+1)\\
\alpha=\lfloor \frac{\sqrt{8i+1}-1}{2}\rfloor\\
\beta=\lfloor \frac{i-s_1+m}{m+1} \rfloor\\
\gamma=\lfloor m-\sqrt{m(m+1)-2(i-s_1-s_2)+9/4}+3/2 \rfloor\\
p=i-s_1-(\beta-1)(m+1)\\
q=i-s_1-s_2-(\gamma-1)(m+1)+\gamma(\gamma-1)/2\\
\end{array}\right.
\eeq
\end{theo}

It requires $O(1)$ time to compute $(-1)^k$ for any positive integer $k$, since

$$
(-1)^k=\left\{\begin{array}{cl}
-1 & \verb"if " k \verb" odd"\\
1 & \verb"if " k \verb" even"\\
\end{array}\right.
$$

Therefore, for each $1\leq i\leq nm+n+m$, $x_i$ can be computed in $O(1)$ time by using the formula (27), and then the optimal move sequence of the general game consisting of $n$ black checkers and $m$ white checkers can be easily computed in optimal $O(nm+n+m)$ time.

\section{The Number of Optimal Solutions}
In this section we will use the state space graph of a game as a tool to discuss the number of optimal solutions of our problem.
A state refers to the status of a game at a given moment. In our problem it must be the positions of the checkers on the checkerboard.
In solving a problem one starts from some initial state and tries to reach a goal state by passing through a series
of intermediate states. In game playing, each move on the game board is a transition from one state to another.
If we think of each state being connected to those states which can follow from it, we have a graph. Such a collection of
interconnected states is called a state space graph.
For example, the initial state and the goal state in our problem are ${\overbrace{b\cdots b}^n} O {\overbrace{w\cdots w}^m}$ and ${\overbrace{w\cdots w}^m} O {\overbrace{b\cdots b}^n}$. A state space graph of the easy case of $n=m=1$ is shown in Fig. 2.

In state based search, a computer program may start from an initial state, then look at one of its successor or children states
and so on until it reaches a goal state. It may reach a dead end state from where it cannot proceed further. In such a situation the program may "backtrack", i.e. undo its last move and try an alternative successor to its previous state.  A path from the initial state to the goal state constitutes a solution. An optimal solution of the problem corresponds to a shortest path from the initial state to the goal state in the state space graph of the problem. Our task in this section is to count the number of different optimal solutions of the problem, which is equivalent to count the number of different shortest paths from the initial state to the goal state in the state space graph of the problem. For example, in the easy case of $n=m=1$, we have two different optimal solutions of the problem, as shown in Fig. 2.

\begin{figure}
\centering
\includegraphics[width=4.5cm,height=4.5cm]{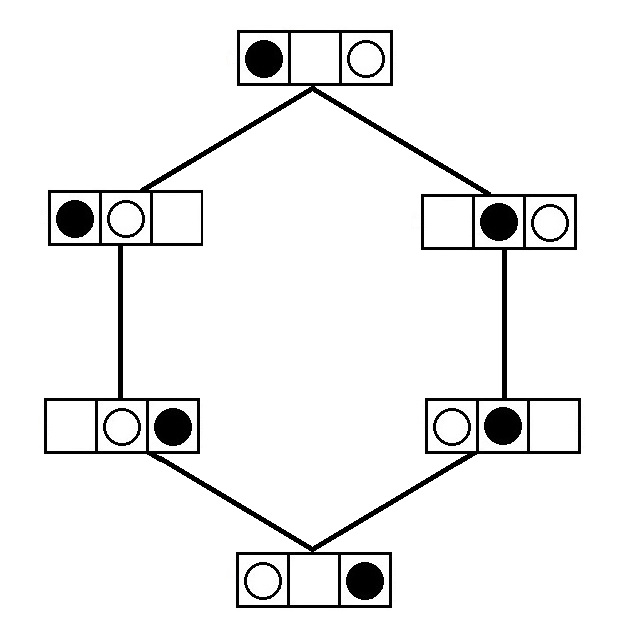}
\caption{A state space graph of the easy case of $n=m=1$}
\end{figure}

In any optimal solutions, the following 4 special states are especially important:
$$\xi_0={\overbrace{b\cdots b}^n} O {\overbrace{w\cdots w}^m}$$
$$\xi_g={\overbrace{w\cdots w}^m} O {\overbrace{b\cdots b}^n}$$
$$\xi_1={\overbrace{b\cdots b}^nw} O {\overbrace{w\cdots w}^{m-1}}$$
$$\xi_2={\overbrace{b\cdots b}^{n-1}} O {b\overbrace{w\cdots w}^m}$$

The state $\xi_0$ is the initial state, and $\xi_g$ is the goal state of the game.
From Lemma 1 we know that in any optimal move sequence, only the classes of moves numbered from 1 to 4 in Table 1 are possible.
With this restriction, our first move from the initial state must be a slide in one direction.
If the first move is $slide(l)$, then the initial state $\xi_0$ will be changed to $\xi_1$.
Otherwise, the first move must be $slide(r)$, and the initial state $\xi_0$ will be changed to $\xi_2$.
In other words, the shortest paths from the initial state $\xi_0$ to the goal state $\xi_g$ must be in the forms $\xi_0,\xi_1,P_1,\xi_g$ or $\xi_0,\xi_2,P_2,\xi_g$, where $\xi_1,P_1,\xi_g$ is a shortest path from $\xi_1$ to the goal state $\xi_g$ and $\xi_2,P_2,\xi_g$ is a shortest path from $\xi_2$ to the goal state $\xi_g$.
If we have made a first move, the following paths $P_1$ or $P_2$ can be determined by the moving rules of Lemma 1.

There are two cases to be distinguished.
\subsection{The general case of $n\geq m>1$}
Without loss of generality, let the first move be $slide(l)$, and the initial state $\xi_0$ be changed to $\xi_1$.
We now consider the shortest path $P_1$.
We have noticed in the proof of Lemma 1 that in any optimal move sequence no two or more pieces of the same color can come together, unless the two ends of the sequence.
From this point of view, we can prove the following facts by induction.
\begin{lemma}\label{le8}
The following special states $\lambda_1,\lambda_2,\cdots,\lambda_m$ must be in the path $P_1$, and for all $i$ such that $1\leq i\leq m-1$, the shortest paths between the states $\lambda_i$ and $\lambda_{i+1}$ are unique.
$$\lambda_1={\overbrace{b\cdots b}^{n-1}} Owb{\overbrace{w\cdots w}^{m-1}}$$
$$\lambda_2={\overbrace{b\cdots b}^{n-2}} wbwbO{\overbrace{w\cdots w}^{m-2}}$$
$$\vdots$$
$$\lambda_t=\left\{\begin{array}{lr}
{\overbrace{b\cdots b}^{n-t}} {\overbrace{wb\cdots wb}^{2t}O{\overbrace{w\cdots w}^{m-t}}} & t\ \mathrm{even}\\
{\overbrace{b\cdots b}^{n-t}}O{\overbrace{wb\cdots wb}^{2t}{\overbrace{w\cdots w}^{m-t}}} & t\ \mathrm{odd}\\
\end{array}\right.$$
$$\vdots$$
$$\lambda_m=\left\{\begin{array}{lr}
{\overbrace{b\cdots b}^{n-m}} {\overbrace{wb\cdots wb}^{2m}O} & m\ \mathrm{even}\\
{\overbrace{b\cdots b}^{n-m}}O{\overbrace{wb\cdots wb}^{2m}} & m\ \mathrm{odd}\\
\end{array}\right.$$
\end{lemma}

\noindent{\bf Proof.}

It is clear that the only way to move optimally from the state $\xi_1$ is a move of $jump(r)$, which changes the state $\xi_1$ to $\lambda_1$.
Similarly, from the state $\lambda_1$, the unique choice among the 4 possible moves is a move of $slide(r)$, otherwise two pieces of the same color will come together, which is forbidden. Following this move, two $jump(l)$ are forced for the same reason. At this point, the state $\lambda_2$ is reached, and the shortest path from $\lambda_1$ to $\lambda_2$ is unique.

Suppose the claim is true for $i<t$. In the case of $i=t<m$, we have to move from the state $\lambda_t$ to $\lambda_{t+1}$.
If $t$ is even, then $\lambda_t=\overbrace{b\cdots b}^{n-t}\overbrace{wb\cdots wb}^{2t}O\overbrace{w\cdots w}^{m-t}$. It is clear that from the state $\lambda_t$, the moves $jump(l)$ and $jump(r)$ will make two pieces of the same color come together. If a $slide(r)$ is applied first, then no jumps are possible in the following moves. The game will reach a dead end to this case. Therefore, the only choice for the first move is $slide(l)$. Following this move, $t+1$ $jump(r)$ are forced for the same reason, and then we reached the state $\lambda_{t+1}$. The shortest path from $\lambda_t$ to $\lambda_{t+1}$ is clearly unique. The claim is therefore true by induction. If $t$ is odd, the proof is similar.

The proof is completed. $\Box$

\begin{lemma}\label{le9}
The following special states $\mu_1,\mu_2,\cdots,\mu_{n-m}$ must be in the path $P_1$, and for all $i$ such that $1\leq i\leq n-m-1$, the shortest paths between the states $\mu_i$ and $\mu_{i+1}$ are unique.

$$\mu_1=\left\{\begin{array}{lr}
{\overbrace{b\cdots b}^{n-m-1}}O{\overbrace{wb\cdots wb}^{2m}b} & m\ \mathrm{even}\\
{\overbrace{b\cdots b}^{n-m-1}}{\overbrace{wb\cdots wb}^{2m}Ob} & m\ \mathrm{odd}\\
\end{array}\right.$$

$$\mu_2=\left\{\begin{array}{lr}
{\overbrace{b\cdots b}^{n-m-2}}{\overbrace{wb\cdots wb}^{2m}Obb} & m\ \mathrm{even}\\
{\overbrace{b\cdots b}^{n-m-2}}O{\overbrace{wb\cdots wb}^{2m}bb} & m\ \mathrm{odd}\\
\end{array}\right.$$
$$\vdots$$
$$\mu_t=\left\{\begin{array}{lr}
{\overbrace{b\cdots b}^{n-m-t}}{\overbrace{wb\cdots wb}^{2m}O\overbrace{b\cdots b}^{t}} & m+t\ \mathrm{even}\\
{\overbrace{b\cdots b}^{n-m-t}}O{\overbrace{wb\cdots wb}^{2m}\overbrace{b\cdots b}^{t}} & m+t\ \mathrm{odd}\\
\end{array}\right.$$
$$\vdots$$
$$\mu_{n-m}=\left\{\begin{array}{lr}
{\overbrace{wb\cdots wb}^{2m}O\overbrace{b\cdots b}^{n-m}} & n\ \mathrm{even}\\
O\overbrace{wb\cdots wb}^{2m}\overbrace{b\cdots b}^{n-m} & n\ \mathrm{odd}\\
\end{array}\right.$$
\end{lemma}
\noindent{\bf Proof.}

If $m$ is even, then $\lambda_m=\overbrace{b\cdots b}^{n-m}\overbrace{wb\cdots wb}^{2m}O$. It is clear that from the state $\lambda_m$, only two moves $jump(r)$ and $slide(r)$ are possible.
If a $jump(r)$ is applied first, then no slides are possible in the following moves. The game will reach a dead end to this case. Therefore, the only choice for the first move is $slide(r)$. Following this move, $m$ $jump(r)$ are forced for the same reason, and then we reached the state $\mu_1$. The shortest path from $\lambda_m$ to $\mu_1$ is clearly unique. In the case of $m$ odd, the analysis is similar.

Suppose the claim is true for $i<t$. For the case of $i=t<n-m$, we have to move from the state $\mu_t$ to $\mu_{t+1}$.

If $m+t$ is odd, then $\mu_t={\overbrace{b\cdots b}^{n-m-t}}O{\overbrace{wb\cdots wb}^{2m}\overbrace{b\cdots b}^{t}}$. It is clear that from the state $\lambda_t$, the moves $jump(l)$ will go back and thus not optimal.
If a $jump(r)$ is applied first, then the game will reach a dead end ${O\overbrace{b\cdots b}^{n-m-t}}{\overbrace{wb\cdots wb}^{2m}\overbrace{b\cdots b}^{t}}$ for this case.
If a $slide(l)$ is applied first, then the next move must be a $jump(r)$, which will lead to the state ${\overbrace{b\cdots b}^{n-m-t-1}}Owbb{\overbrace{wb\cdots wb}^{2m-2}\overbrace{b\cdots b}^{t}}$. In this state, the two black pieces come together, and they are not at the right end since $m>1$. This is impossible.
Therefore, the only choice for the first move is $slide(r)$. Following this move, $m$ $jump(l)$ are forced for the same reason, and then we reached the state $\mu_{t+1}$. The shortest path from $\mu_t$ to $\mu_{t+1}$ is clearly unique. The claim is therefore true by induction.

If $m+t$ is even, then $\mu_t={\overbrace{b\cdots b}^{n-m-t}}{\overbrace{wb\cdots wb}^{2m}O\overbrace{b\cdots b}^{t}}$. It is clear that from the state $\lambda_t$, the moves $jump(l)$ and $jump(r)$ will go back and thus not optimal.
If a $slide(l)$ is applied first, then the game will reach a dead end ${\overbrace{b\cdots b}^{n-m-t}}{\overbrace{wb\cdots wb}^{2m}\overbrace{b\cdots b}^{t}}O$ for this case.
Therefore, the only choice for the first move is $slide(r)$. Following this move, $m$ $jump(l)$ are forced for the same reason, and then we reached the state $\mu_{t+1}$. The shortest path from $\mu_t$ to $\mu_{t+1}$ is clearly unique. The claim is therefore true by induction.

The proof is completed. $\Box$

\begin{lemma}\label{le10}
The following special states $\nu_1,\nu_2,\cdots,\nu_m$ must be in the path $P_1$, and for all $i$ such that $1\leq i\leq m-1$, the shortest paths between the states $\nu_i$ and $\nu_{i+1}$ are unique.

$$\nu_1=\left\{\begin{array}{lr}
wO\overbrace{wb\cdots wb}^{2(m-1)}\overbrace{b\cdots b}^{n-m+1} & n\ \mathrm{even}\\
{w\overbrace{wb\cdots wb}^{2(m-1)}O\overbrace{b\cdots b}^{n-m+1}} & n\ \mathrm{odd}\\
\end{array}\right.$$

$$\nu_2=\left\{\begin{array}{lr}
{ww\overbrace{wb\cdots wb}^{2(m-2)}O\overbrace{b\cdots b}^{n-m+2}} & n\ \mathrm{even}\\
wwO\overbrace{wb\cdots wb}^{2(m-2)}\overbrace{b\cdots b}^{n-m+2} & n\ \mathrm{odd}\\
\end{array}\right.$$
$$\vdots$$
$$\nu_t=\left\{\begin{array}{lr}
{\overbrace{w\cdots w}^{t}\overbrace{wb\cdots wb}^{2(m-t)}O\overbrace{b\cdots b}^{n-m+t}} & n+t\ \mathrm{even}\\
\overbrace{w\cdots w}^{t}O\overbrace{wb\cdots wb}^{2(m-t)}\overbrace{b\cdots b}^{n-m+t} & n+t\ \mathrm{odd}\\
\end{array}\right.$$
$$\vdots$$
$$\nu_m=\overbrace{w\cdots w}^{m}O\overbrace{b\cdots b}^{n}$$
\end{lemma}
\noindent{\bf Proof.}

The claim of this lemma is symmetric to Lemma 8, and therefore the proof is also symmetric.$\Box$

Combining Lemma 8,9 and 10, we conclude that in the shortest path $\xi_0,\xi_1,P_1,\xi_g$ from the initial state $\xi_0$ to the goal state $\xi_g$, the shortest path $P_1$ must be unique.
The analysis for the case of first move $slide(r)$ is similar, and we can conclude also that in the shortest path $\xi_0,\xi_2,P_2,\xi_g$ from the initial state $\xi_0$ to the goal state $\xi_g$, the shortest path $P_2$ must be unique.
Finally we conclude that in the general case of $n\geq m>1$, there are only two different shortest paths from the initial state $\xi_0$ to the goal state $\xi_g$. Therefore, in this case, the number of optimal solutions of the game is 2. The two different optimal solutions of the game can be computed by the formula \eqn{eq27} of Theorem~\ref{th3} in linear time.

\subsection{The special case of $n\geq m=1$}
In this special case, Lemma 8 and 10 are also true. Therefore, the shortest paths from the initial state $\xi_0$ to the state $\lambda_1$, and the shortest path from the state $\mu_{n-1}$ to the state $\nu_1=\xi_g$ are still unique.
The special states of Lemma 9 become complicated in the case of $m=1$, since the shortest paths between any two consecutive states of these special states are no longer unique. In this special case, we will expand the special states $\mu_1, \mu_2, \cdots, \mu_{n-m}$ of Lemma~\ref{le9} further to $\mu_{ij},0\leq i\leq n-1, 1\leq j\leq 2$ as follows.
$$\left\{\begin{array}{l}
\mu_{01}=\overbrace{b\cdots b}^{n-1}wbO\\
\mu_{02}=\overbrace{b\cdots b}^{n-1}Owb
\end{array}\right.$$
$$\left\{\begin{array}{l}
\mu_{11}=\overbrace{b\cdots b}^{n-2}wbOb\\
\mu_{12}=\overbrace{b\cdots b}^{n-2}Owbb
\end{array}\right.$$
$$\vdots$$
$$\left\{\begin{array}{l}
\mu_{t1}=\overbrace{b\cdots b}^{n-t-1}wbO\overbrace{b\cdots b}^{t}\\
\mu_{t2}=\overbrace{b\cdots b}^{n-t-1}Owb\overbrace{b\cdots b}^{t}
\end{array}\right.$$
$$\vdots$$
$$\left\{\begin{array}{l}
\mu_{(n-1)1}=wbO\overbrace{b\cdots b}^{n-1}\\
\mu_{(n-1)2}=Owb\overbrace{b\cdots b}^{n-1}
\end{array}\right.$$
The claim of Lemma~\ref{le9} will modified to the following Lemma~\ref{le11}.
\begin{lemma}\label{le11}
The special states $\mu_{ij},1\leq i\leq n-1, 1\leq j\leq 2$ must be in the path $P_1$ or $P_2$. For each $i$ such that $0\leq i\leq n-2$, there is only one shortest path from the state $\mu_{i1}$ to the state $\mu_{(i+1)2}$; there are two shortest paths from the state $\mu_{i2}$, one to the state $\mu_{(i+1)1}$, and the other to the state $\mu_{(i+1)2}$.
\end{lemma}
\noindent{\bf Proof.}

It is clear that $\mu_{01}=\xi_1$ and $\mu_{02}=\xi_2$. Two moves $slide(r)$ and $jump(r)$ change the state $\mu_{01}$ to $\mu_{12}$. Two moves $slide(r)$ and $jump(l)$ change the state $\mu_{02}$ to $\mu_{11}$, and another two moves $slide(l)$ and $jump(r)$ change the state $\mu_{02}$ to $\mu_{12}$.

For the general case of $0\leq i\leq n-m-1$, two moves $slide(r)$ and $jump(r)$ change the state $\mu_{i1}$ to $\mu_{(i+1)2}$;
two moves $slide(r)$ and $jump(l)$ change the state $\mu_{i2}$ to $\mu_{(i+1)1}$, and another two moves $slide(l)$ and $jump(r)$ change the state $\mu_{i2}$ to $\mu_{(i+1)2}$.

Notice that the length of the shortest paths from $\mu_{01}=\xi_1$ and $\mu_{02}=\xi_2$ to $\mu_{(n-1)1}$ and $\mu_{(n-1)2}$ is $2(n-1)$ by Lemma 4, we conclude that The special states $\mu_{ij},1\leq i\leq n-1, 1\leq j\leq 2$ must be in the path $P_1$ or $P_2$.

The proof is completed. $\Box$

Denote the number of different shortest paths from the state $\mu_{ij}$ to the states $\mu_{(n-1)1}$ or $\mu_{(n-1)2}$ as $\rho(i,j)$, then from Lemma~\ref{le11} we have,
\beql{eq29}
\left\{\begin{array}{l}
\rho(i,1)=\rho(i+1,2)\\
\rho(i,2)=\rho(i+1,1)+\rho(i+1,2)\\
\rho(n-2,1)=1\\
\rho(n-2,2)=2
\end{array}\right.
\eeq

The solution of this recurrence is

\beql{eq30}
\left\{\begin{array}{l}
\rho(i,1)=F_{n-i}\\
\rho(i,2)=F_{n-i+1}
\end{array}\right.
\eeq

where $F_n$ is the $n$th Fibonacci number
$\frac{1}{\sqrt{5}}\left( \left(\frac{1+\sqrt{5}}{2}\right)^n-\left(\frac{1-\sqrt{5}}{2}\right)^n\right)$.

Therefore, in this case, the number of different shortest paths from the initial state $\xi_0$ to the goal state $\xi_g$ is
$\rho(0,1)+\rho(0,2)=F_n+F_{n+1}=F_{n+2}$.

Summing up, the number of optimal solutions for solving the general game of shifting the checkers consisting of $n$ black checkers and $m$ white checkers can be given in the following Theorem.

\begin{theo}\label{th4}
For the general game of shifting the checkers consisting of $n$ black checkers and $m$ white checkers, let $\varphi(n,m)$ be the number of optimal solutions for solving the game, then $\varphi(n,m)$ can be expressed explicitly as follows.
\beql{eq31}
\varphi(n,m)=\left\{\begin{array}{cr}
\frac{1}{\sqrt{5}}\left( \left(\frac{1+\sqrt{5}}{2}\right)^{n+2}-\left(\frac{1-\sqrt{5}}{2}\right)^{n+2}\right) & n\geq m=1\\
2 & n\geq m>1
\end{array}\right.
\eeq
\end{theo}

\section{Concluding Remarks}
We have studied the general shifting the checkers game consisting of $n$ black checkers and $m$ white checkers. It has been proved in the section 2 that the minimum number of steps needed to play the game for general $n$ and $m$ is $nm + n + m$. All of the optimal solutions for the moving checkers game of small size can be found by a backtracking algorithm presented in section 2.
In the section 3, a linear time recursive construction algorithm which can produce an optimal solution in linear time for very large size $n$ and $m$ is presented. The time cost of the new algorithm is $O(nm)$ and $O(n+m)$ space is used.
In Section 4, an extremely simple explicit solution for the optimal moving sequences of the general game is given. The formula gives for each individual step $i$, its optimal move in $O(1)$ time. Finally, in Section 5 we give the complete optimal solutions for the game in general cases.

Another similar game is to reverse the $n$ checkers numbered $1,\cdots, n$ by two permissible types of moves $slide$ and $jump$. It is not clear whether our methods presented in this paper can be applied to this game. We will investigate the problem further.


\begin{thebibliography}{99}
\bibitem{1}
R. Bird, Pearls of Functional Algorithm Design, 258-274, Cambridge University Press, 2010.

\bibitem{2}
Erik D. Demaine, Playing games with algorithms, Algorithmic combinatorial game theory. Proceedings of the 26th Symposium on Mathematical Foundations in Computer Science, LNCS 2136, 18-32, 2001.

\bibitem{3}
Erik D. Demaine and Martin L. Demaine, Puzzles, Art, and Magic with Algorithms, Theory of Computing Systems, vol. 39, number 3, 473-481, 2006.

\bibitem{4}
A. Levitin and M. Levitin, Algorithmic Puzzles, 3-31, Oxford University Press, New York, 2011.

\bibitem{5}
J. Kleinberg, E. Tardos. Algorithm Design, 223-238, Addison Wesley, 2005.

\bibitem{6}
D.L. Kreher and D. Stinson, Combinatorial Algorithms: Generation, Enumeration and Search, 125-133, CRC Press, 1998.

\bibitem{6}
John S. Gray, The shuttle puzzle ¡ª A lesson in problem solving, Journal of Computing in Higher Education, Volume 10, Issue 1, 56-70, 1998.

\end{thebibliography}
\end{document}